\documentclass[conference]{IEEEtran}
\IEEEoverridecommandlockouts
\usepackage[T1]{fontenc}
\usepackage[utf8]{inputenc}
\usepackage{textcomp}
\usepackage{graphicx}
\usepackage{amsmath,amssymb}
\usepackage{booktabs}
\usepackage{array}
\usepackage{microtype}
\usepackage{url}
\usepackage{cite}

\usepackage[hidelinks]{hyperref}
\usepackage[export]{adjustbox}
\usepackage{placeins}

\title{Void-Free Through-Glass Vias for Multilayer Surface-Electrode Ion Traps}
\author{
\IEEEauthorblockN{
Eike Iseke\IEEEauthorrefmark{1},
Jacob Stupp\IEEEauthorrefmark{2},
Friederike Giebel\IEEEauthorrefmark{1},
Nila Krishnakumar\IEEEauthorrefmark{1},
Konstantin Thronberens\IEEEauthorrefmark{1},\\
Nora D. Stahr\IEEEauthorrefmark{2},
Md. Masum Billah\IEEEauthorrefmark{2},
Marlon Kuhn\IEEEauthorrefmark{2},
Christian Ospelkaus\IEEEauthorrefmark{1}\IEEEauthorrefmark{2}
}
\IEEEauthorblockA{
\IEEEauthorrefmark{1}Physikalisch-Technische Bundesanstalt (PTB), Germany\\
\IEEEauthorrefmark{2}Gottfried Wilhelm Leibniz Universität Hannover, Germany
}
\thanks{
© 2026 IEEE.  Personal use of this material is permitted.  Permission from IEEE must be obtained for all other uses, in any current or future media, including reprinting/republishing this material for advertising or promotional purposes, creating new collective works, for resale or redistribution to servers or lists, or reuse of any copyrighted component of this work in other works.
}
}

\begin{document}
\maketitle

\begin{abstract}
Through-substrate interconnects are a key technology for scaling surface-electrode ion traps (SEITs) toward large electrode counts and complex architectures. Dielectric substrates such as fused silica offer superior RF performance compared to silicon, but the reliable fabrication of void-free through-glass vias (TGVs) remains challenging. In this work, we demonstrate a SEIT fabricated on a fused silica substrate incorporating TGVs. The via-holes are formed using Selective Laser-Induced Etching and subsequently metallized with gold by electroplating. A patent-pending geometry-assisted bottom-up plating approach enables void-free filling using constant-current electroplating without electrolyte additives or pulsed deposition schemes. Electrical characterization across multiple wafers confirms a TGV-level yield above 98.8 \%, demonstrating the robustness and reproducibility of the developed TGV technology. The TGVs are combined with multilayer SEIT fabrication, enabling vertical signal routing to all electrodes while maintaining an unaffected trapping surface without wire bonds. To interface the trap with external control electronics, a glass interposer providing signal fan-out to a larger pitch is designed and fabricated. The trap chip and the interposer are integrated using thermocompression bonding, forming a compact chip-to-chip assembly. The presented technology demonstrates the first fully integrated multilayer SEIT on fused silica incorporating void-free TGVs.
\end{abstract}


\section{Introduction}
The efficient simulation of quantum mechanical systems on classical computers is fundamentally limited by the exponential growth of the Hilbert space with system size \cite{r1}. This insight laid the conceptual foundation for quantum computing, where controllable quantum systems are used to process information beyond classical capabilities. Among the most promising hardware platforms are trapped ions. They provide a scalable architecture in which laser-mediated coupling between internal electronic states and collective motional modes enables universal quantum gate operations \cite{r2}. In this context, surface-electrode ion traps (SEITs) have emerged as a leading platform for scalable quantum computing \cite{r3}. Their planar design provides a robust framework for implementing the shuttling-based Quantum Charge-Coupled Device (QCCD) architecture \cite{r4,r5,r6}. SEIT-based processors have recently been deployed in commercial quantum computing systems \cite{r7,r8}, highlighting the technological maturity and practical viability of this platform.

Beyond laser-driven gates, the close proximity of ions to microfabricated electrodes in SEITs has allowed microwave-based control to emerge as an alternative approach for coherent quantum logic with trapped ions. Entangling gates generated by integrated near-field microwave conductors in microfabricated ion traps have been demonstrated \cite{r9}. In addition, microwave field gradients enable individual-ion addressing \cite{r10}. By eliminating the need for complex laser beam delivery systems, microwave-driven gate operations substantially reduce optical overhead and simplify system architecture. This approach facilitates a scalable and highly integrable hardware platform \cite{r9,r10}.

Within this platform, our group developed near-field microwaves \cite{r11,r12,r13}, implemented a two-qubit microwave gate \cite{r14}, introduced an amplitude modulation technique that enabled high-fidelity two-qubit gates \cite{r15,r16} and realized a universal two-qubit processor \cite{r17}. An 8-qubit multilayer SEIT with chip-integrated microwave control was subsequently designed and fabricated \cite{r18}.

In general, for the trapped-ion platform to increase electrode density and overcome routing limitations of single-layer geometries, advanced multilayer fabrication processes have been developed \cite{r19,r20}. These processes enable complex electrode layouts and integrated control wiring within a compact footprint.
Despite these advancements, the external electrical connection to high-density electrode arrays remains a bottleneck. Conventionally, electrodes are contacted via wire bonding at the chip periphery \cite{r21}. This method constrains the layout because contact pads must be located at the chip edges, which can consume significant chip area. Furthermore, wire bonds can physically obstruct the low-angle optical access \cite{r21} required for laser cooling and state detection. To circumvent these limitations, through-substrate interconnects such as through-silicon vias (TSVs) or through-glass vias (TGVs) provide a promising approach. By routing signals to the backside of the substrate, they increase interconnect density and decouple wiring from the trapping surface. This enables more scalable device architectures. Initial demonstrations of trapped-ion devices incorporating through-substrate interconnects have been reported. These implementations used silicon substrates \cite{r22,r23}. However, experiments with glass- or quartz-like substrates show higher Q-factors, lower insertion loss, and overall superior RF performance compared to Si-based traps \cite{r24,r25,r26}. To combine the advantages of through-substrate interconnects with the favorable RF properties of dielectric substrates, a reliable and process-stable TGV technology is required. The fabrication of high-aspect-ratio TGVs with void-free metallization remains challenging \cite{r27,r28}, as conventional conformal metallization approaches often suffer from premature closure and void formation.

In this work, we address this challenge by introducing a geometry-assisted metallization approach for TGVs in fused silica. The via-holes are fabricated using Selective Laser-Induced Etching and filled with gold by electroplating. A tailored geometry enables controlled bottom-up growth and allows void-free metallization using simple constant-current plating without electrolyte additives or pulsed deposition schemes. Based on this patent-pending technology, we demonstrate the first multilayer SEIT fabricated on a fused silica substrate incorporating TGVs. Recent work has demonstrated chiplet-based architectures with dielectric substrates and through-substrate vias, providing key building blocks for scalable systems \cite{r29}. Here we realize a fully integrated device combining multilayer electrodes on the front side and contact pads connected through the TGVs on the backside. This enables vertical signal routing while maintaining an unobstructed trapping surface. To interface the trap with external control electronics, a glass interposer providing signal fan-out to a larger pitch is designed and fabricated. The ion trap chip and the interposer are finally connected using thermocompression bonding, forming a compact chip-to-chip assembly.

Together, the presented TGV technology, multilayer trap fabrication, interposer design, and chip integration establish a scalable platform for large-scale glass-based SEITs.
\section{Through-Glass Vias}
Reliable TGVs are required to provide low-resistance electrical interconnects while maintaining compatibility with multilayer trap fabrication. The through-via structures must ensure continuous metallization, mechanical stability, and void-free filling to prevent electrical failure and long-term reliability issues. Furthermore, a void-free metallization prevents the formation of virtual leaks in the vacuum system during operation of the trap. This section describes the formation of high-aspect-ratio through-vias in fused silica, the limitations observed with conformal metallization, and the development of a geometry-assisted bottom-up electroplating process enabling void-free gold filling.
\begin{figure}[t]
    \centering
    \includegraphics[width=0.95\columnwidth]{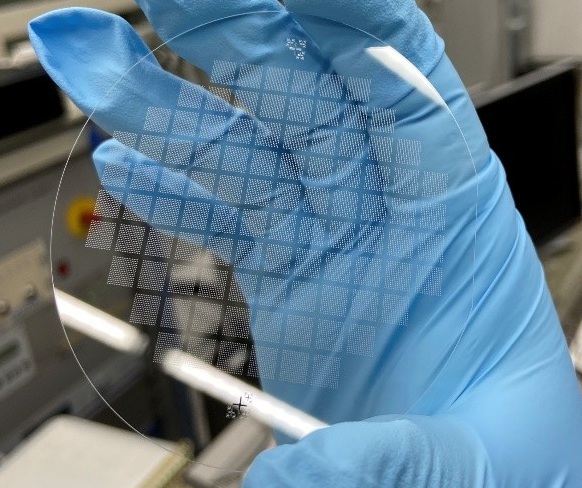}
    \caption{3-inch fused silica wafer after Selective Laser-Induced Etching (SLE). The wafer contains 88 chips, each comprising 220 via-holes with a diameter of 75~$\mu$m.}
    \label{fig:1}
\end{figure}
\subsection{Selective Laser-Induced Etching}
Conventional TSVs in silicon are typically fabricated using deep reactive ion etching, also known as the Bosch process \cite{r30,r31}. For glass substrates, laser-based structuring techniques such as laser-induced deep etching (LIDE) \cite{r32} and Selective Laser-Induced Etching (SLE) \cite{r33} are available. While LIDE enables high-throughput via-hole formation, SLE allows greater flexibility in shaping complex three-dimensional geometries. We therefore employ SLE to realize the required via-hole structures in fused silica.

SLE is a two-step process consisting of laser modification and subsequent wet chemical etching. First, the via-hole geometry is defined digitally and sliced into multiple layers with a step size of 5~$\mu$m. For a substrate thickness of 380~$\mu$m, the resulting structure is divided into 77 layers. A femtosecond laser scans each layer and locally modifies the glass volume. The ultrashort pulses induce structural changes in the material, including formation of low-density void-like regions \cite{r34}. After laser exposure, the wafer is wet-chemically etched in a KOH solution. While the entire substrate is exposed to the etchant, the laser-modified regions dissolve significantly faster than the unmodified material. Reported etch selectivities reach values of up to 1400 \cite{r35} and 3000 \cite{r36}. This strong differential etching enables the removal of the written volumes and results in the formation of via-holes that can subsequently be metallized to realize TGVs.

Initial via-holes were fabricated with straight sidewalls. This geometry was intended for conformal metallization. The metallization results are shown in Section II.B. Subsequently, shaped via-holes were realized to enable geometry-assisted bottom-up filling. The results of metallizing these structures are shown in Section II.C. The shaping is achieved by controlled laser exposure of the sidewall regions and does not require any additional steps as compared to the straight via-holes.

The SLE process was performed on 3-inch fused silica wafers with a thickness of 380~$\mu$m. Each wafer contains 88 chips with a size of 5~mm~$\times$~5~mm. Each chip comprises 220 via-holes with a pitch of 300~$\mu$m and a diameter of 75~$\mu$m. The resulting via-holes exhibit an aspect ratio of approximately 5. After etching, all via-holes were fully opened. Across the complete wafer, corresponding to 19360 via-holes, a via-hole formation yield of 100 \% was achieved. Fig.~\ref{fig:1} shows a picture of a wafer after the SLE process.
\subsection{Conformal Metallization of straight via-holes}\begin{figure}[t]
    \centering
    \includegraphics[width=0.95\columnwidth]{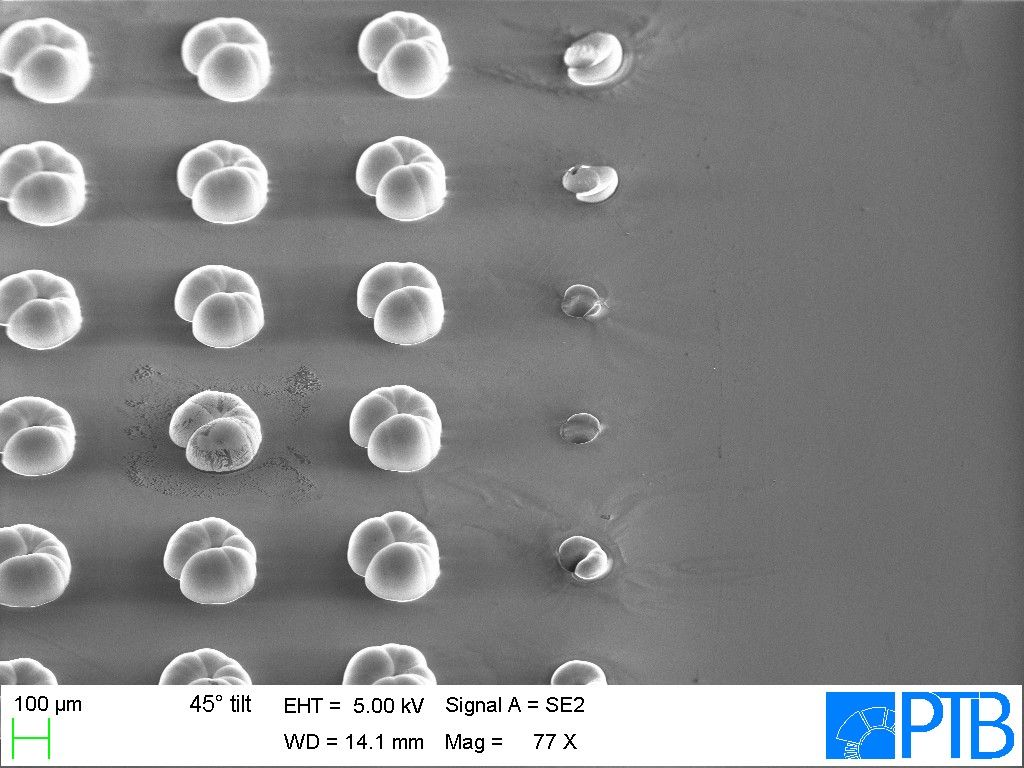}
    \caption{SEM image of neighboring TGVs after conformal electroplating. Most TGVs exhibit pronounced overgrowth at the TGV entrance, while several remain only partially filled.}
    \label{fig:2}
\end{figure}
\begin{figure}[b]
    \centering
    \includegraphics[width=0.95\columnwidth]{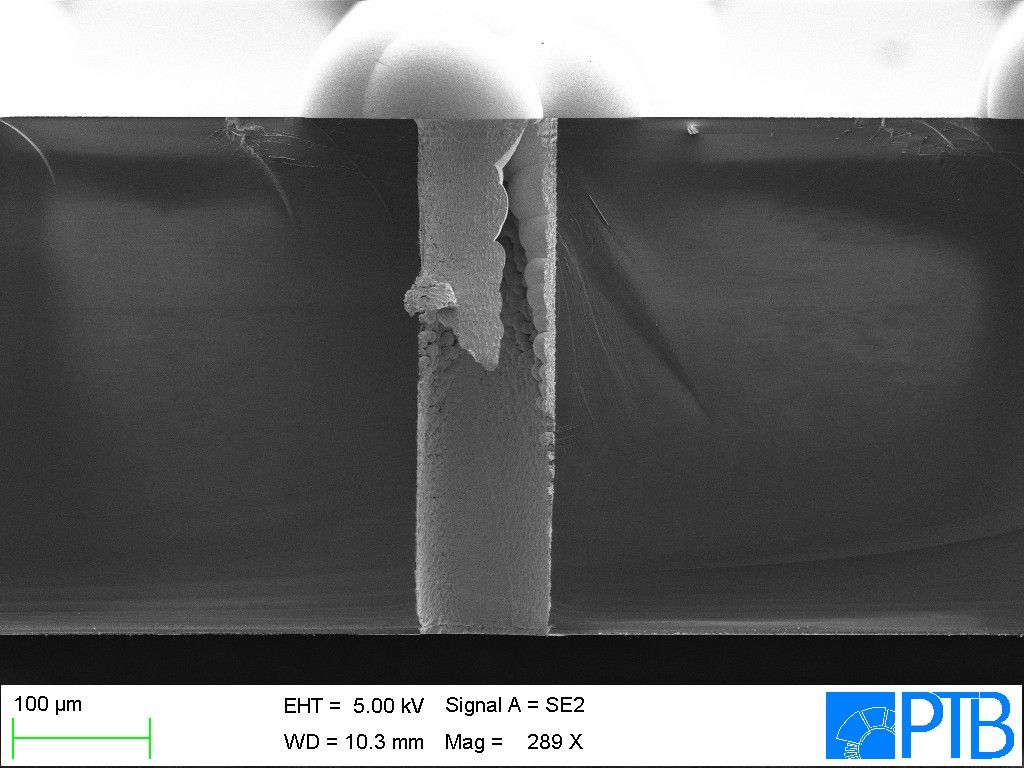}
    \caption{Cross-sectional SEM image of a TGV after conformal electroplating. Premature closure at the via-hole entrance is observed while the lower region remains mostly unfilled, resulting in a large internal void.}
    \label{fig:3}
\end{figure}
Given the straight via-hole geometry described above, a conformal metallization approach was initially pursued to minimize plating time. In this scheme, metal growth proceeds from the sidewalls toward the center, requiring a maximum growth distance of 37.5~$\mu$m instead of 380~$\mu$m for bottom-up filling.

Prior to electroplating, a conductive seed stack is deposited. It consists of 17 nm titanium followed by 50 nm gold. Titanium improves the adhesion of gold to fused silica, while the gold itself provides a conductive starting layer for electroplating. The seed stack is deposited by sputtering. The wafer holder is positioned off-axis relative to the target, and the wafer is rotated during deposition. This configuration results in coating of the bottom side of the wafer and conformal coating of the via-hole sidewalls.

Electrical contact for electroplating is established on the coated bottom side of the wafer. After electrically contacting the wafer, the gold coated side is sealed to prevent deposition on the wafer surface. Electroplating was performed under constant current conditions.

The resulting filling behavior is strongly inhomogeneous. An overview SEM image of neighboring TGVs is shown in Fig.~\ref{fig:2}. Most TGVs exhibit pronounced overgrowth at the top surface, while several remain only partially filled. A representative cross section is shown in Fig.~\ref{fig:3}. The via-hole entrance closes early during plating, while the lower region remains largely unfilled, resulting in the formation of a sizeable internal void.

This behavior is attributed to an effect known as current crowding \cite{r37,r38}. An accumulation of field lines at the via-hole rim leads to increased local current density and therefore higher local deposition rates \cite{r31}. Metal growth at the via-hole entrance therefore outpaces deposition at the bottom, causing premature closure and void formation.

\begin{figure}[t]
    \centering
     \includegraphics[width=0.95\columnwidth]{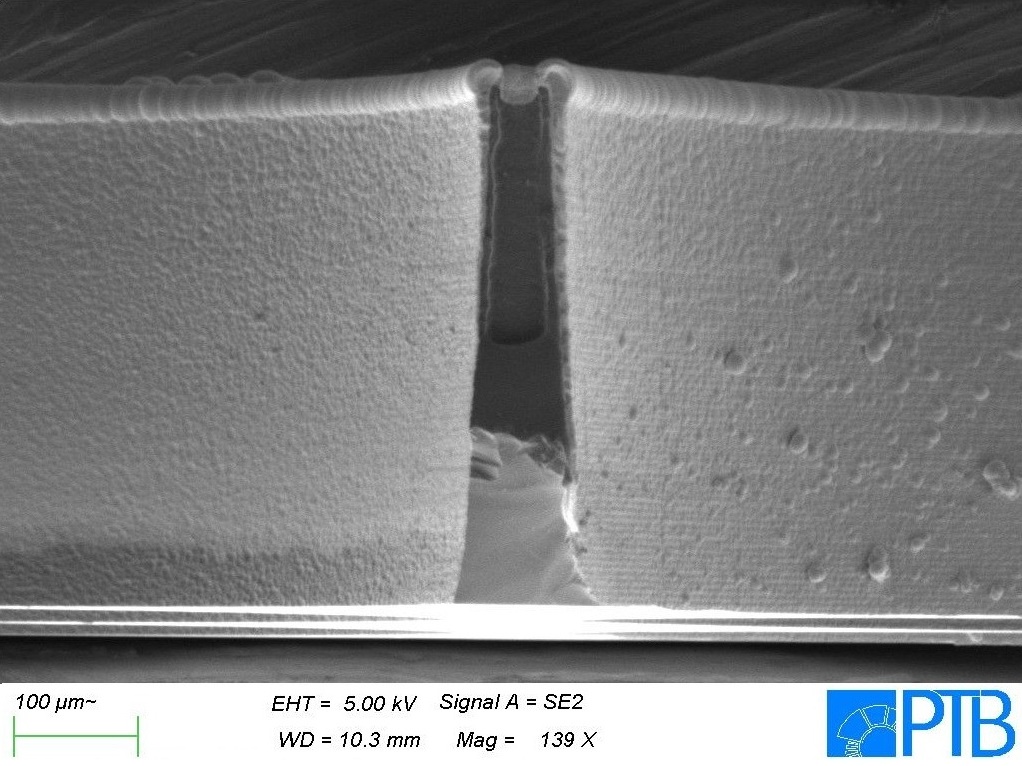}
    \caption{Cross-sectional SEM image of two adjacent test TGVs with a diameter of 1~mm after electroplating. Gold deposition occurs along the sidewalls, while enhanced growth at the via-hole rim (top part of the picture) leads to pronounced overgrowth caused by current crowding.}
    \label{fig:4}
\end{figure}

To verify this mechanism, test structures with enlarged via-hole diameters of 1~mm were fabricated and electroplated to a thickness of approximately 10~$\mu$m. The cross section shown in Fig.~\ref{fig:4} reveals enhanced edge growth compared to the central region, confirming the current-density-driven deposition inhomogeneity.

Conformal direct current metallization of straight high-aspect-ratio via-holes without chemical additives therefore does not result in reliable void-free filling.
\subsection{Geometry-Assisted Bottom-Up Metallization}
Although void-free filling of high-aspect-ratio via-holes has been demonstrated in other application domains, these approaches typically rely on tailored electrolyte additives \cite{r39,r40,r41} or pulsed and reverse-pulsed plating \cite{r42,r43,r44} schemes. Such process control increases complexity and reduces process robustness. To achieve reliable void-free filling using a simple and stable process flow, a geometry-driven solution was developed. The plating behavior is controlled solely by engineering the via-hole geometry, eliminating the need for complex additives or time-dependent deposition schemes.
\begin{figure}[t]
    \centering
    \includegraphics[width=0.95\columnwidth]{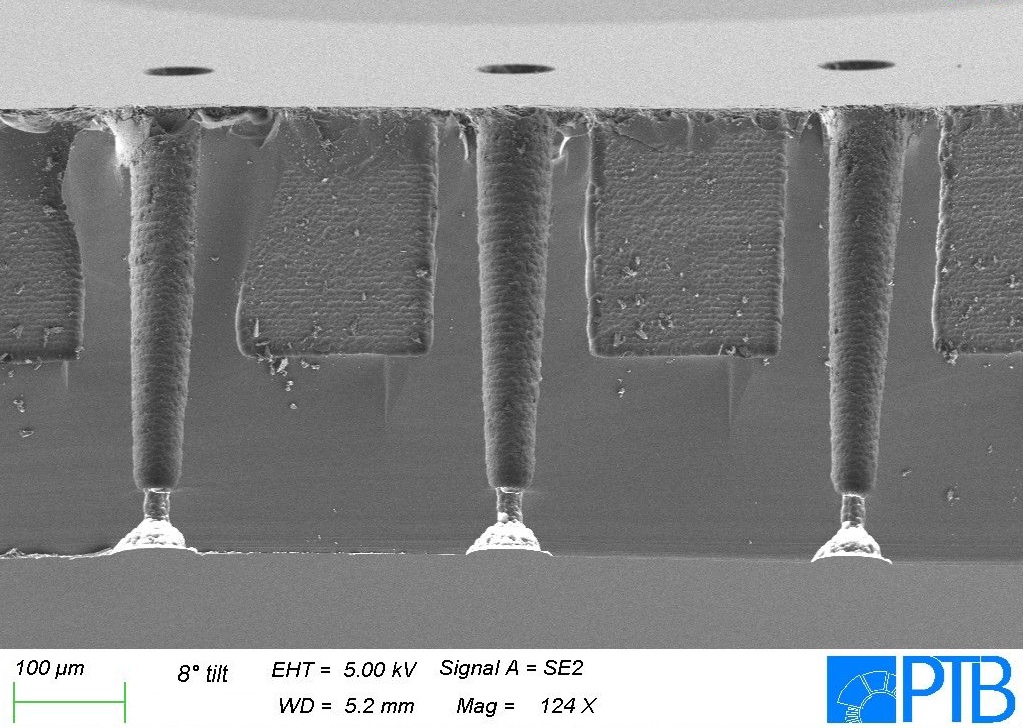}
    \caption{Cross-sectional SEM image of three neighboring via-holes after seed layer deposition. Bright contrast in the chamfer region indicates the presence of the sputtered gold seed layer, while the upper via-hole region remains uncoated. Additional notches introduced by SLE enable controlled wafer cleaving for cross-section analysis.}
    \label{fig:5}
\end{figure}
\begin{figure} [b]
    \centering
    \includegraphics[width=0.95\columnwidth]{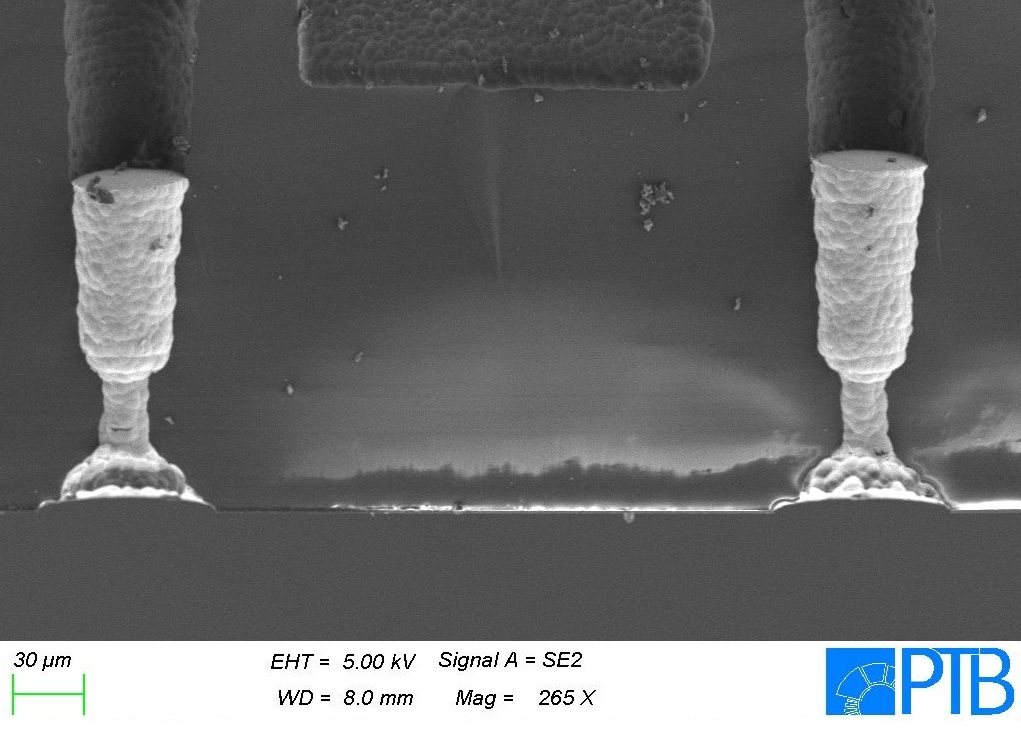}
    \caption{Cross-sectional SEM image of two neighboring TGVs after approximately 100~$\mu$m of electroplated gold. Deposition initiates in the lower region and proceeds uniformly in both structures, demonstrating controlled bottom-up filling without void formation.}
    \label{fig:6}
\end{figure}

The via-hole geometry was modified to enable controlled bottom-up growth. The via-hole consists of a 45$^\circ$ chamfer on the backside, followed by a narrow tunnel section. The chamfer and the tunnel prevent deposition of the sputtered seed layer into the upper via-hole region. After the tunnel, the structure opens again at 45$^\circ$ before transitioning into the main via-hole with a 3$^\circ$ taper.

Because of the described geometry, after seed layer deposition, gold is present only in the lower chamfer region and on the backside surface. The tunnel as well as the main via-hole remain uncoated. An SEM image of three adjacent via-holes after seed deposition is shown in Fig.~\ref{fig:5}. The bright contrast observed in the chamfer region indicates the presence of the sputtered gold seed layer. The absence of similar contrast in the rest of the via-hole confirms that the seed layer is confined to the chamfer.

During electroplating, deposition starts exclusively in the lower region of the via-holes. Current crowding at the chamfer edges promotes rapid lateral growth. The tunnel closes first, establishing a continuous metal front across the via-hole cross section. After closure of the tunnel, growth proceeds vertically toward the front side of the wafer, resulting in true bottom-up filling.

A cross section of two neighboring TGVs after plating of approximately 100~$\mu$m of gold is shown in Fig.~\ref{fig:6}. Both TGVs exhibit identical filling heights, indicating uniform growth across the wafer. No void formation is observed.

\begin{figure}[t]
    \centering
    \includegraphics[width=0.95\columnwidth,  trim=0 0 0 39pt,
    clip]{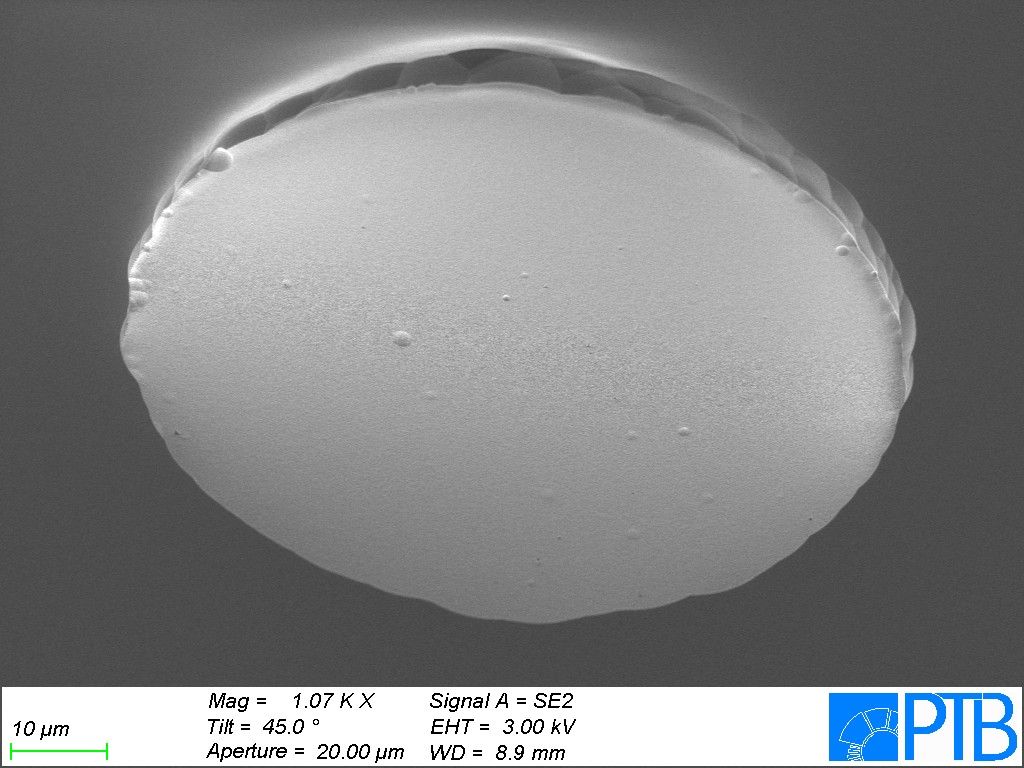}
    \caption{SEM image of a completely filled TGV on the front side of the wafer. The gold surface is flat and homogeneous and terminates slightly below the substrate surface, providing a suitable starting point for subsequent multilayer fabrication.}
    \label{fig:7}
\end{figure}

After complete filling, the front side shows a flat and continuous gold surface, as shown in Fig.~\ref{fig:7}. This surface is suitable for subsequent multilayer processing. On the backside, a shallow indentation remains due to lateral growth within the chamfer before tunnel closure. This does not affect device integration, as large backside contact pads are formed in a subsequent step.

\begin{figure}
    \centering
    \includegraphics[width=0.95\columnwidth]{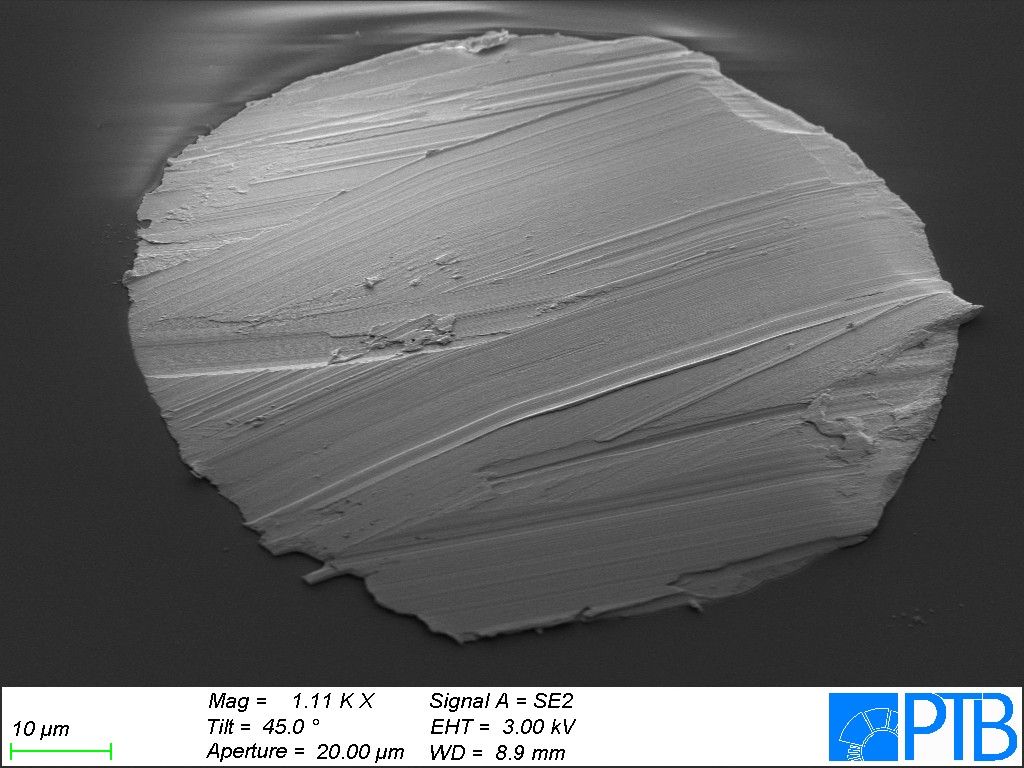}
    \caption{SEM image of an overgrown TGV after mechanical planarization. Excess gold is removed to obtain a surface flush with the substrate, enabling further multilayer processing.}
    \label{fig:8}
\end{figure}

In some cases, slight overgrowth above the TGV opening occurs. Excess gold is mechanically removed after plating. Fig.~\ref{fig:8} shows an overgrown TGV after planarization.
The demonstrated filling behavior enables reliable metallization of the TGVs which are electrically characterized in the following section.
\FloatBarrier
\subsection{Electrical Characterization and Yield}
To assess the reliability of the bottom-up metallization process, electrical measurements were performed on three wafers at room temperature. For each chip, 141 out of 220 TGVs were electrically characterized using an automated probing system. The remaining TGVs are dedicated to ground and thermal management and were not included in the resistance scan.

TGVs with a resistance exceeding 25~$\Omega$ were classified as defective. The TGV yield is defined as the fraction of functional TGVs relative to all measured TGVs. The chip yield is defined as the fraction of chips for which all 141 measured TGVs meet the resistance criterion. A single defective TGV therefore results in a failing chip.

The resistance statistics and yield values for all three wafers are summarized in Table~\ref{tab:results}. 
\begin{table}
\centering
\small
\setlength{\tabcolsep}{2pt}
\caption{Electrical characterization results of three TGV wafers including mean resistance, median resistance, standard deviation, TGV yield, and chip yield.}
\label{tab:results}
\begin{tabular}{c c c c c c}
\toprule
Wafer & Mean ($\Omega$) & Median ($\Omega$) & Std. dev. ($\Omega$) & TGV yield & Chip yield \\
\midrule
1 & 6.62 & 6.52 & 1.31 & 99.81~\% & 85.23~\% \\
2 & 6.67 & 6.46 & 1.62 & 99.24~\% & 62.5~\% \\
3 & 6.64 & 6.48 & 1.65 & 98.85~\% & 62.5~\% \\
\bottomrule
\end{tabular}
\end{table}The measured resistance includes the contact resistance of the two-point probing setup and therefore does not represent the intrinsic resistance of the TGVs. The mean TGV resistance is 6.62~$\Omega$, 6.67~$\Omega$, and 6.64~$\Omega$ for wafers 1, 2, and 3, respectively. The corresponding medians are 6.52~$\Omega$, 6.46~$\Omega$, and 6.48~$\Omega$. The standard deviation ranges from 1.31~$\Omega$ to 1.65~$\Omega$.

The TGV yield amounts to 99.81~\%, 99.24~\%, and 98.85~\%, while the resulting chip yield is 85.23~\%, 62.5~\%, and 62.5~\%, respectively. The close agreement of the resistance statistics and yield values across all wafers indicates stable and reproducible process conditions across multiple wafer runs.

Overall, the electrical characterization confirms high TGV-level reliability and demonstrates the robustness of the geometry-assisted bottom-up metallization process over multiple wafers. These results provide the foundation for subsequent multilayer ion trap fabrication which is described in the following section.
\section{Device Fabrication and Packaging}
Based on the demonstrated TGV technology, a multilayer SEIT was fabricated on the fused silica substrate. The device integrates front-side multilayer electrodes with backside contact pads interconnected through the TGVs. This architecture enables vertical signal routing while maintaining an unobstructed trapping surface without wire bonds.

To provide a scalable electrical interface, the ion trap chip is combined with a dedicated interposer. The interposer redistributes the electrical signals from the trap to a larger pitch and routes them through TGVs to the backside of the interposer, where the electrical connection to the surrounding electronics can be established through a filterboard or a socket.

The following sections describe the layout of the ion trap and the interposer, the respective fabrication processes, and the final chip integration using thermocompression bonding.
\subsection{Ion trap design}
\begin{figure}
    \centering
    \includegraphics[width=0.95\columnwidth]{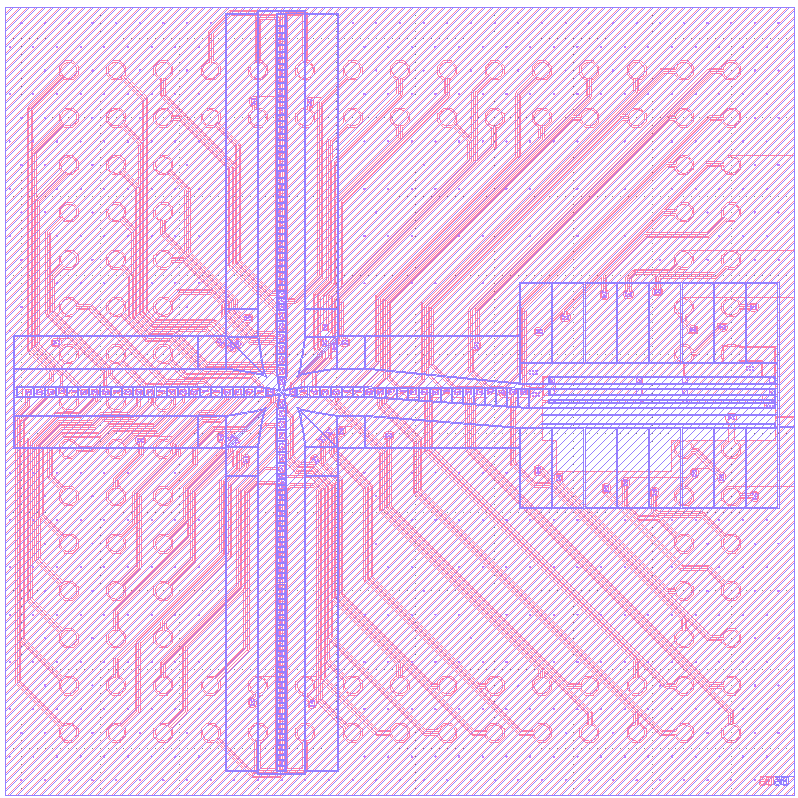}
    \caption{Layout of the ion trap electrode structure. The routing layer is shown in red, while the via layer and top electrode layer are shown in blue. The design follows an X-junction architecture enabling ion transport between four linear trap segments.}
    \label{fig:9}
\end{figure}

The ion trap chip has a lateral size of 5~mm~$\times$~5~mm and incorporates 220 TGVs. The trap follows an X-junction architecture that enables ion transport between four linear segments. Such junction geometries allow scalable trap operations by shuttling ions between different functional zones. These regions can be used for ion loading, transport, storage, and quantum gate operations. The overall electrode layout is shown in Fig.~\ref{fig:9}.

The electrode structure is realized using a multilayer metallization scheme. The upper metal layer defines the trapping electrodes and forms the SEIT. A lower metal layer is used for signal routing across the chip. Both layers are connected through an intermediate layer consisting of dielectric insulation and vertical interconnects. This multilayer architecture allows dense routing of control signals without increasing the footprint of the trap electrodes.

The trap includes one RF electrode for radial confinement, one microwave conductor for qubit control, and 110 DC electrodes for axial confinement and ion transport. Each electrode is connected to a single TGV. The combination of multilayer routing and TGVs enables vertical signal access to all electrodes.

On the backside of the chip, rectangular contact pads are formed around the TGV openings. These pads provide the electrical interface for connecting the ion trap chip to an interposer using thermocompression bonding.
\subsection{Interposer Design}
The interposer provides the electrical interface between the ion trap chip and the external control electronics. It redistributes the signals from the trap chip to a larger pitch and routes them through the substrate to the backside, where electrical connections to the surrounding electronics can be established. The interposer has a lateral size of 40~mm~$\times$~40~mm and is fabricated on fused silica wafers with a diameter of 3 inches.

The signal routing is implemented on the front side of the interposer using a single metal layer. The electrical signals from the ion trap chip are first routed laterally across the interposer and subsequently transferred through the substrate using TGVs. On the backside of the interposer chip, rectangular contact pads surrounding the TGV openings provide the electrical interface to the external electronics.

To increase robustness against individual TGV failures, redundant TGVs are implemented for each signal line. Each signal is routed through four parallel TGVs, which connect the front-side routing layer to the backside contact pads. This redundancy increases the robustness of the interposer against individual TGV defects.
\subsection{Ion Trap Fabrication}
The ion trap chips are fabricated on 3-inch fused silica wafers with a thickness of 380~$\mu$m after completion of the TGV process described in Section II. Prior to further processing, residual gold overgrowth originating from the bottom-up TGV metallization is mechanically removed. This step produces a smooth front surface suitable for subsequent lithographic processing. At this stage the electrical integrity of all TGVs has already been verified, as discussed in Section II. The fabrication sequence consists of backside contact pad formation followed by multilayer electrode fabrication on the front side of the wafer.

Due to the via-hole geometry used for bottom-up electroplating, a shallow crater remains around the TGV openings on the backside of the wafer. The backside metallization is performed first, since the titanium/gold seed layer stack deposited during the TGV metallization process is already present on this side of the wafer and can be directly reused for the fabrication of the backside pads.

A lithography step using a positive photoresist and UV mask aligner defines the contact pad geometry. After exposure and development, the wafer is electrically contacted through the seed layer and gold is electroplated to form the backside pads. The pads are grown to a thickness of approximately 2~$\mu$m. Following electroplating, the photoresist is removed in acetone using an ultrasonic bath.

\begin{figure} [b]
    \centering
    \includegraphics[width=0.95\columnwidth]{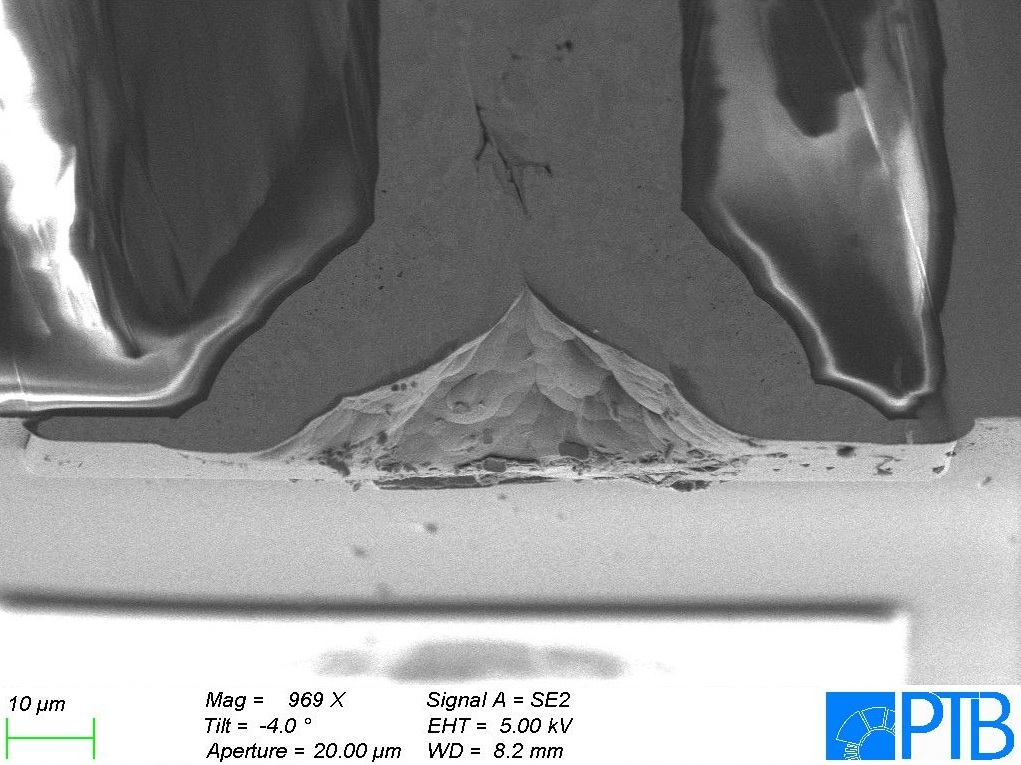}
    \caption{Cross-sectional SEM image of the lower part of a TGV with the electroplated backside contact pad. The characteristic crater shape around the TGV opening originates from the bottom-up metallization process.}
    \label{fig:10}
\end{figure}

The then exposed seed layer stack is subsequently removed by reactive ion etching. The etch is performed without a lithographic mask and therefore removes a thin gold layer uniformly across the wafer surface. Since the electroplated pads are significantly thicker than the seed layer, the removal of 50 nm of gold does not affect the pad geometry. After this step the backside metallization is complete. A cross-section of the bottom part of a TGV together with the backside pad is shown in Fig.~\ref{fig:10}. The cross-section was prepared using argon ion milling. The image reveals the characteristic crater shape around the TGV opening as well as the electroplated contact pad.

\begin{figure}[t]
    \centering
    \includegraphics[width=0.95\columnwidth]{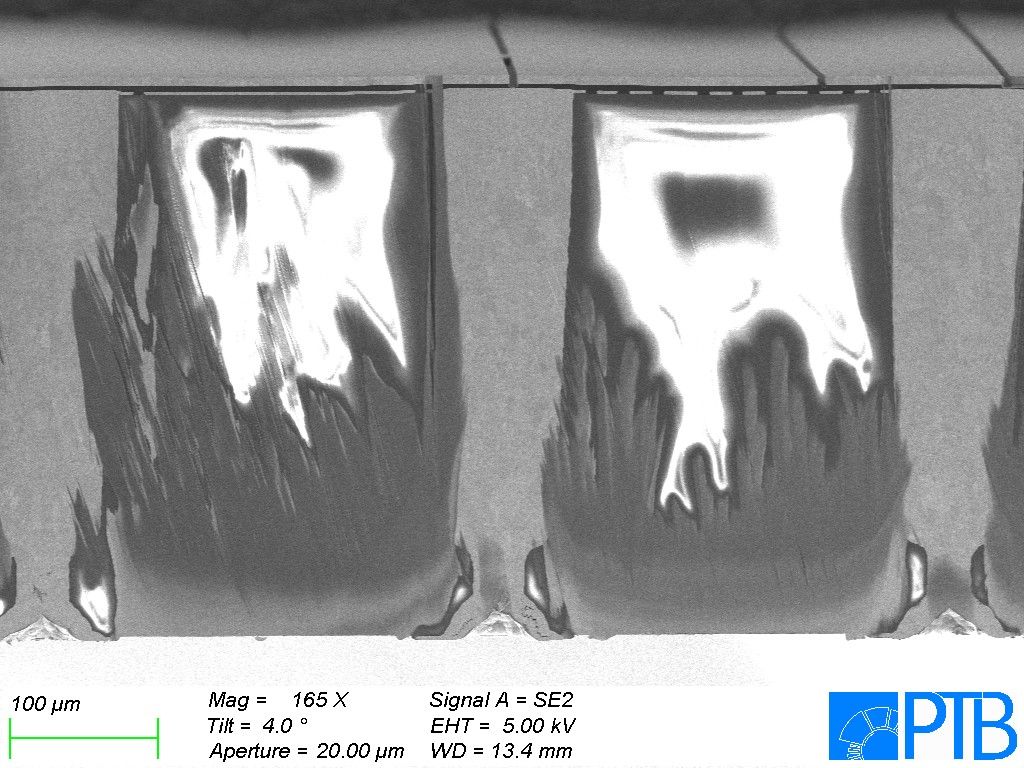}
    \caption{Cross-sectional SEM image of the complete multilayer structure (top part of the picture) including the TGV, routing layer, multilayer vias, and top electrode layer. The TGVs are fully metallized. Bright regions correspond to the fused silica substrate and result from dielectric charging during SEM imaging.}
    \label{fig:11}
\end{figure}
\begin{figure} [b]
    \centering
    \includegraphics[width=0.95\columnwidth]{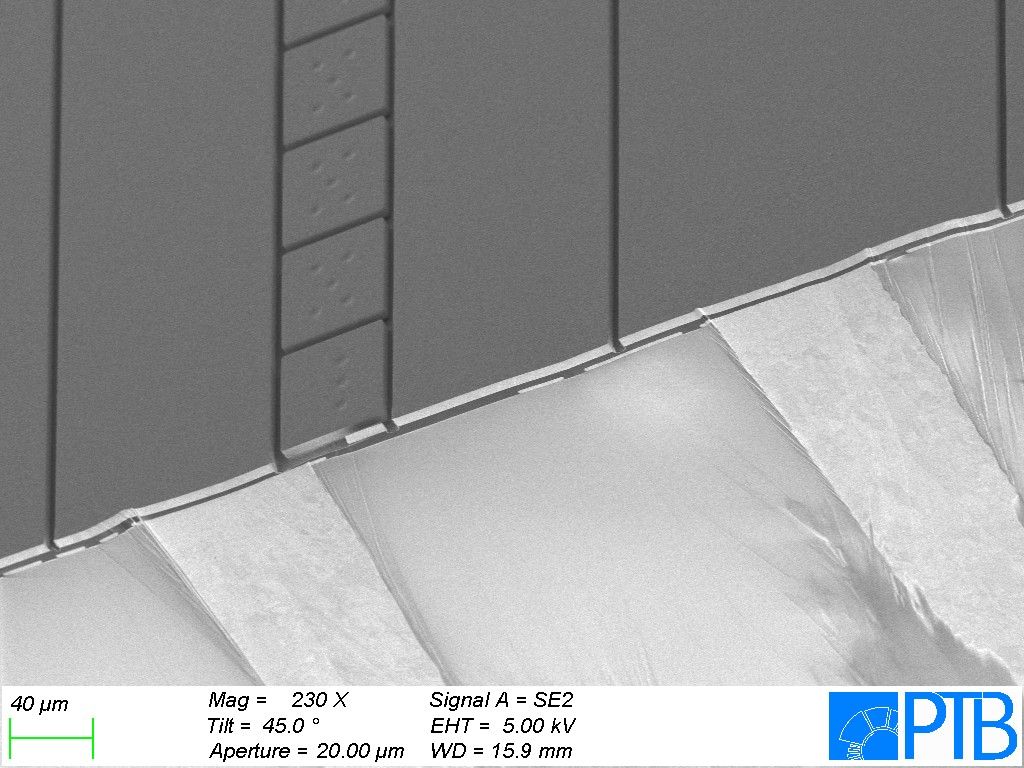}
    \caption{Tilted cross-sectional SEM image showing two neighboring TGVs with the multilayer electrode stack on top. Multilayer vias connecting the routing layer to the top electrode layer are visible.}
    \label{fig:12}
\end{figure}

After completion of the backside processing, fabrication continues on the front side of the wafer where the multilayer electrode structure of the ion trap is realized. The multilayer process follows the general approach described in previous work of our group \cite{r19} and enables complex electrode routing on the trap chip.

The fabrication begins with the deposition of a titanium/gold seed layer (17 nm titanium / 50 nm gold) by physical vapor deposition. This layer serves as the seed layer for subsequent electroplating steps. A lithography step defines the routing layer structures. Gold is then electroplated to a thickness of approximately 2~$\mu$m to form the first metal layer. This routing layer distributes electrical signals from the TGVs to the electrode locations defined in the upper trap layer.

After electroplating, the photoresist is removed. The wafer then undergoes a second lithography step defining the multilayer vias that link the routing layer to the upper electrode layer. These vias have a diameter of approximately 7.5~$\mu$m and are electroplated to a thickness of about 3~$\mu$m. The previously deposited seed layer is reused for this electroplating step. After plating, the resist is removed and the remaining seed layer is etched.

In the next step an insulating dielectric layer is applied to electrically isolate the routing layer from the top electrode layer. A polyimide film is spin coated onto the wafer to a thickness of approximately 6~$\mu$m in order to fully cover the routing and via layer. The polyimide is subsequently cured at 250$^\circ$C to obtain the final dielectric properties.

During spin coating the topography of the underlying metal structures is transferred into the polyimide layer. To achieve a planar surface and to expose the multilayer vias, a chemical mechanical polishing (CMP) step is performed. The CMP process removes polyimide until the tops of the electroplated vias are reached while leaving the metal structures intact.

To ensure reliable electrical contact to the vias, an additional lithography and etching step is carried out after planarization. Photoresist is applied everywhere except on the via openings and a brief etch removes any residual polyimide remaining on the via surfaces. This step guarantees that the multilayer vias are completely exposed before deposition of the top electrode layer.

The fabrication of the final electrode layer then proceeds in a manner similar to the first metal layer. A titanium/gold seed layer is deposited by physical vapor deposition. Lithography defines the electrode geometry of the ion trap including RF, microwave, and DC electrodes. Gold is electroplated to a thickness of approximately 6~$\mu$m to form the final electrode structures.

\begin{figure}[t]
    \centering
    \includegraphics[width=0.95\columnwidth]{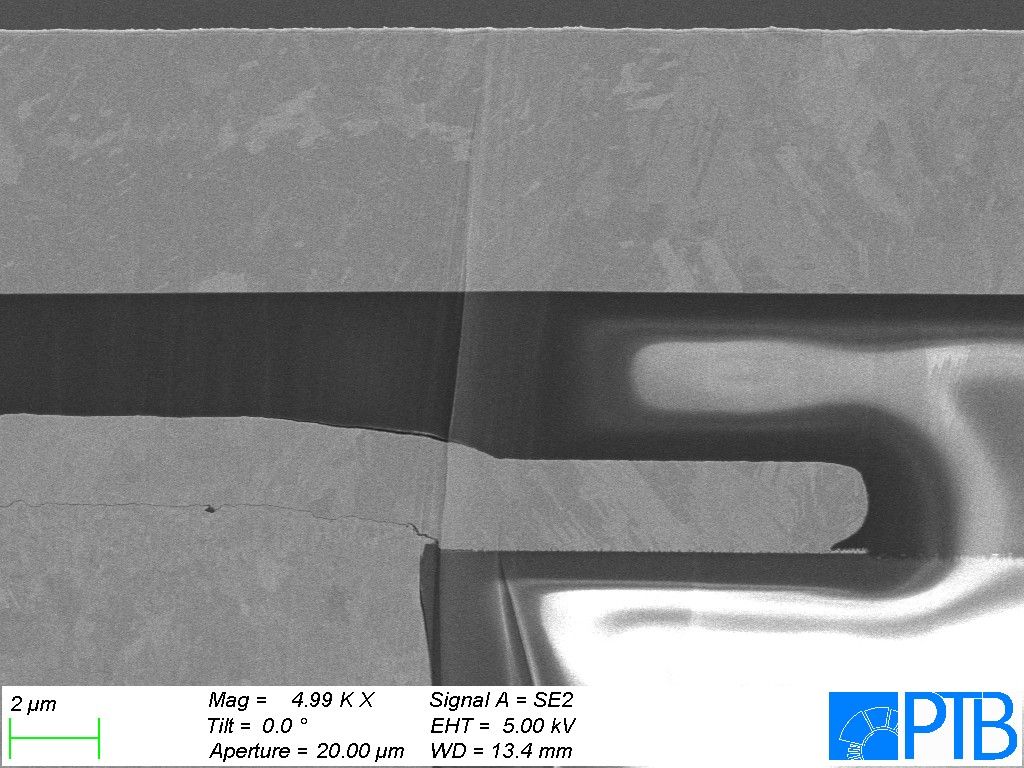}
    \caption{High-magnification cross-section of the multilayer stack. The routing layer lies directly on the substrate (in the middle) and on top of a TGV (on the left side). Planarization of the polyimide dielectric by CMP compensates the topography difference, enabling deposition of a flat top electrode layer.}
    \label{fig:13}
\end{figure}

After electroplating the photoresist is removed and the exposed seed layer is etched. This completes the fabrication of the multilayer electrode stack consisting of the routing layer, the vertical interconnect layer, and the top electrode layer. A cross-section of the complete structure is shown in Fig.~\ref{fig:11}. The image illustrates the TGV, the routing layer, the interconnect vias, and the final electrode layer. The cross-section shows the successful integration of the TGVs with the multilayer electrode stack. The bright areas between the TGVs correspond to the glass substrate and result from charging of the dielectric material during SEM imaging. Fig.~\ref{fig:12} shows a more tilted view of the cross section. Two neighboring TGVs are visible, although only their upper sections are shown. The multilayer electrode stack can again be observed on top of the TGVs. Small circular features are visible on the inner DC electrodes of the top metal layer. These correspond to the vias of the multilayer structure. They appear due to a slight topography difference introduced during the via opening process. Fig.~\ref{fig:13} shows a higher magnification cross-section of the multilayer stack. In the middle of the image, the first metal layer lies directly on the fused silica substrate, while on the left side of the image the layer lies on top of the TGV. The TGV protrudes slightly above the surrounding substrate surface by approximately 1~$\mu$m, which introduces a small topography difference in the first metal layer. During the subsequent process steps, the polyimide layer is deposited and planarized by CMP. As a result, the final electrode layer is deposited on a planar surface and does not inherit the underlying topography. This demonstrates that the CMP process effectively compensates the height difference introduced by the TGVs and confirms the compatibility of the multilayer process with the integrated TGV structures.

Finally, the processed wafers are diced into individual ion trap chips using a conventional wafer saw. Each chip has a lateral size of 5~mm~$\times$~5~mm and contains the complete multilayer electrode structure interconnected through the TGVs.

The described fabrication sequence combines TGV technology with multilayer SEIT fabrication on a fused silica substrate. To the best of our knowledge, this represents the first demonstration of a multilayer SEIT on a dielectric glass substrate incorporating TGVs.
\subsection{Interposer Fabrication}
The interposer chips are fabricated on 3-inch fused silica wafers using a process flow similar to the ion trap fabrication described above. The TGVs are first created using the same SLE and bottom-up electroplating process introduced in Section II. After TGV fabrication, all TGVs are electrically tested to ensure reliable vertical interconnections across the wafer. Only wafers with fully functional TGVs are processed further.

\begin{figure}[t]
    \centering
    \includegraphics[width=0.95\columnwidth, trim=0 20pt 0 50pt, clip]{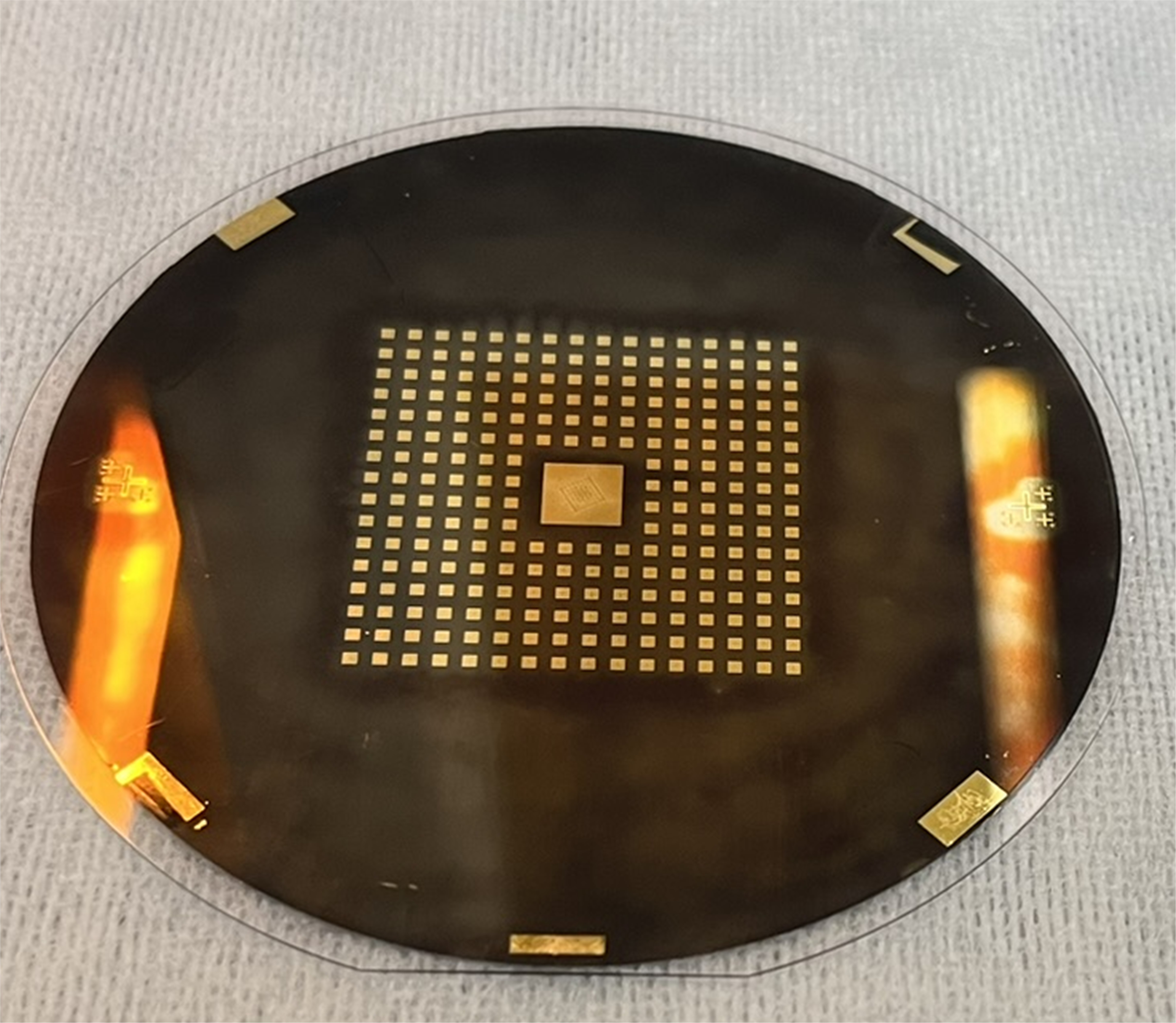}
    \caption{Optical image of the backside of the fused silica interposer wafer after electroplating of the contact pads. The large central pad is used for thermal management, while the surrounding pads provide electrical connections to external control electronics.}
    \label{fig:14}
\end{figure}

The backside contact pads are fabricated next. The Ti/Au seed layer used during TGV metallization is reused as the seed layer for electroplating. A lithography step defines the pad structures, followed by gold electroplating to form the backside pads. After resist removal, the exposed seed layer is removed by reactive ion etching. Fig.~\ref{fig:14} shows an optical image of the backside of the interposer wafer with the electroplated contact pads. The larger central pad is used for cooling of the ion trap assembly, while the surrounding pads provide the electrical interface to the external control electronics.

\begin{figure}[b]
    \centering
    \includegraphics[width=0.95\columnwidth]{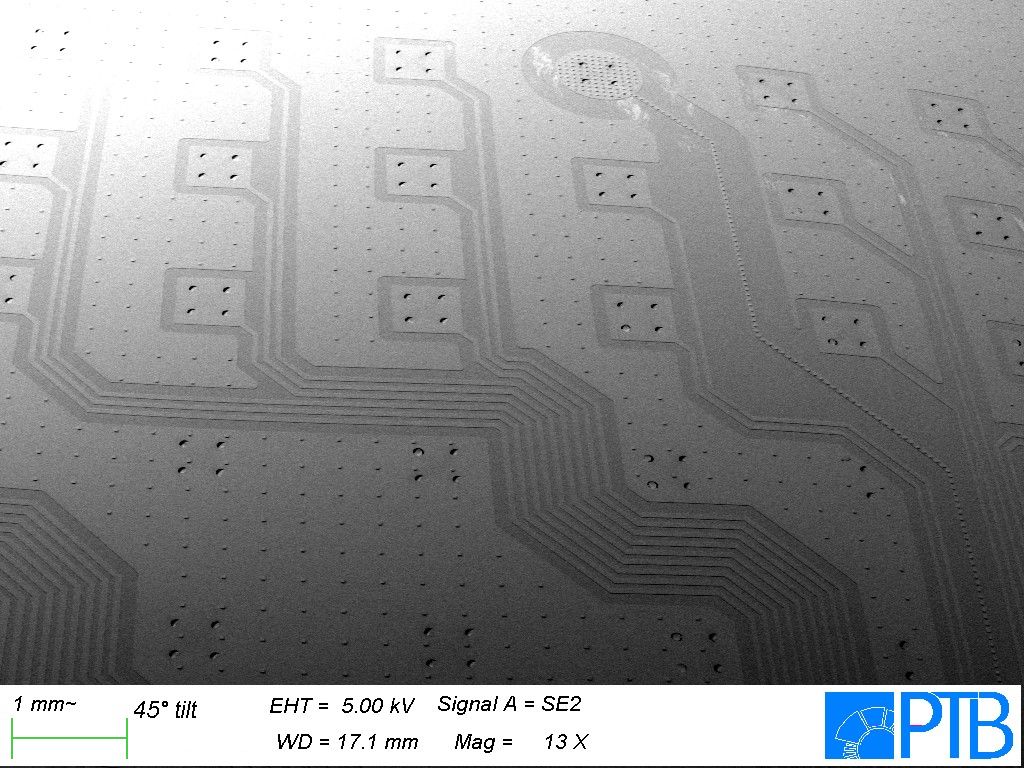}
    \caption{SEM image of a section of the interposer routing layer. The signal lines extend from the central bonding region toward the outer contact pads.}
    \label{fig:15}
\end{figure}

After completion of the backside processing, the front-side routing layer is fabricated. The routing layer is realized using a single electroplated gold layer patterned by UV lithography and electroplating. This layer distributes the signals from the bonding region of the ion trap chip toward the outer contact pads of the interposer. Fig.~\ref{fig:15} shows an SEM image of a section of the routing layer. The image illustrates how the signal lines extend from the central bonding region toward the outer contact pads.

Most bonding pads are designed with a square geometry. A single circular pad is implemented to serve as an alignment marker during the thermocompression bonding process.
\FloatBarrier
\begin{figure}[t]
    \centering
    \includegraphics[width=0.95\columnwidth]{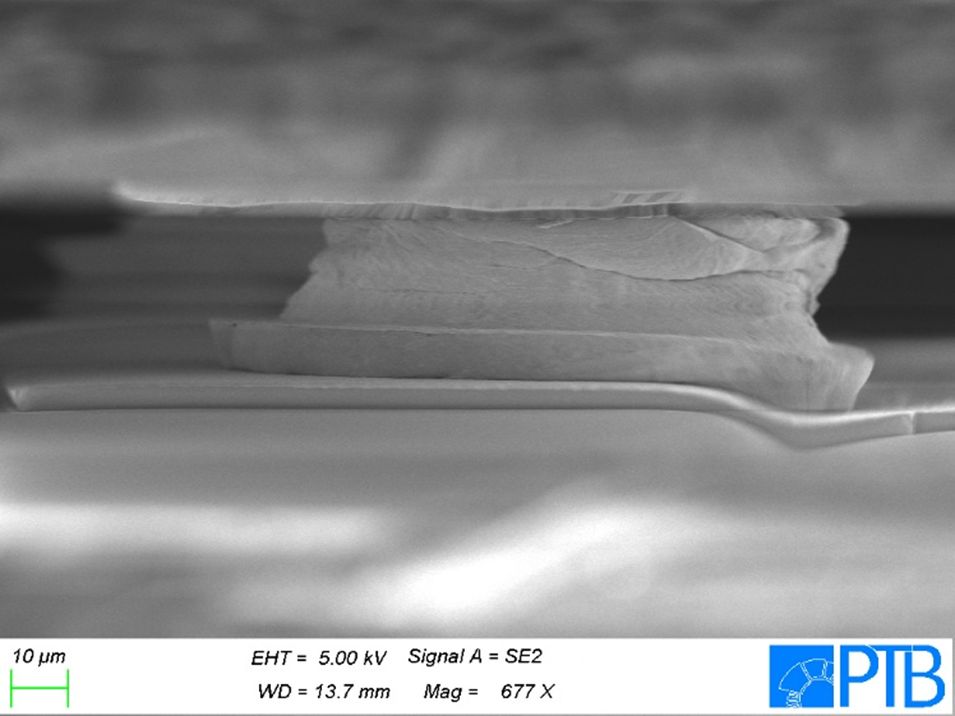}
    \caption{SEM image of a bonded interconnect between the ion trap chip and the interposer. The thermocompressed gold bump forms the electrical connection between the backside pad of the trap chip and the corresponding interposer pad.}
    \label{fig:16}
\end{figure}
\subsection{Thermocompression Bonding}
Electrical and mechanical integration of the ion trap chip with the interposer is achieved using thermocompression bonding. Gold stud bumps are first formed on the backside contact pads of the ion trap chip using a standard wire bonder. The bumps are positioned directly above the TGVs and fit into the shallow craters surrounding the TGV openings created during the TGV fabrication process.

During assembly the ion trap chip is aligned with the corresponding bonding pads on the interposer. Bonding is performed using a fineplacer system from the company Finetech.
Under the combined influence of temperature and pressure, the gold bumps form a reliable electrical and mechanical connection between the ion trap chip and the interposer pads. This thermocompression bonding process has previously been validated for similar device structures \cite{r45} and enables robust chip-to-chip interconnection without the need for additional solder materials.

Fig.~\ref{fig:16} shows a SEM image of a representative bonded interconnect from the side. The image reveals the thermocompressed gold bump that bridges the gap between the two chips, forming the electrical contact between the backside pad of the ion trap chip and the corresponding interposer pad.

The thermocompression bonding approach provides a compact and reliable method to electrically interface the TGV-based ion trap chip with the larger-scale routing structure of the interposer.
\section{Conclusion}
In this work, the first multilayer SEIT incorporating TGVs in a fused silica substrate was realized together with a dedicated glass interposer for scalable signal fan-out. High-aspect-ratio via-holes were fabricated using SLE and metallized by electroplating. A patent-pending geometry-assisted bottom-up plating approach enabled reliable void-free filling using simple constant-current electroplating without electrolyte additives or pulsed plating schemes.

Electrical characterization across multiple wafers confirmed a TGV-level yield above 98.8 \%, demonstrating the robustness and reproducibility of the developed TGV technology. The TGVs were combined with multilayer SEIT fabrication directly on the glass substrate. Vertical signal routing through the TGVs provides electrical access to all electrodes while maintaining an unobstructed trapping surface without wire bonds.

A glass interposer was designed and fabricated to redistribute the trap signals to a larger pitch. The ion trap chip and the interposer were integrated using thermocompression bonding, forming a compact chip-to-chip assembly.

Future work will focus on experimental validation of the trap, including ion loading, ion shuttling experiments and microwave-driven quantum gate operations. In addition, RF performance and microwave loss of the glass-based trap architecture will be investigated.

The successful integration of void-free TGVs with multilayer ion trap fabrication on fused silica establishes a scalable platform for future large-scale trapped-ion quantum processors.
\section*{Acknowledgment}
We express our gratitude and acknowledge the support of the PTB and LNQE cleanroom centers. We acknowledge funding from the BMFTR through the ATIQ, MIQRO, and QVLS-iLabs projects, from the State of Lower Saxony and the Volkswagen Foundation through Quantum Valley Lower Saxony Q1 (QVLS-Q1), and from the Deutsche Forschungsgemeinschaft (DFG, German Research Foundation) under Germany’s Excellence Strategy – EXC-2123 QuantumFrontiers.

\end{document}